%% file: root.tex
\documentclass{ifacconf}

\usepackage{graphicx}      
\usepackage{natbib}        
\counterwithin*{section}{part}

\input{settings/packages}
\input{settings/definitions}
\input{settings/abbreviations}
\makeatletter
\renewcommand{\paragraph}{%
  \@startsection{paragraph}{4}{\z@}%
  {1.0ex \@plus 0.5ex \@minus .2ex}
  {-1em}
  {\normalfont\normalsize\it}
}
\makeatother
\begin{document}

\begin{frontmatter}

\title{%
Smooth Sampling-Based Model Predictive Control Using Deterministic Samples%
} 

\thanks[footnoteinfo]{%
This work is part of the German Research Foundation (DFG) AI Research Unit 5339 regarding the combination of physics-based simulation with AI-based methodologies for the fast maturation of manufacturing processes.%
}%

\author[First]{Markus Walker}
\author[First]{Marcel Reith-Braun}
\author[Second]{Tai Hoang}
\author[Second]{Gerhard Neumann}
\author[First]{Uwe D. Hanebeck}

 \address[First]{Intelligent Sensor-Actuator-Systems Laboratory (ISAS), 
 Institute for Anthropomatics and Robotics, 
 Karlsruhe Institute of Technology, Germany, 
    \{%
        firstname%
    \}%
    .%
    \{%
      lastname%
    \}
    {@kit.edu}%
}
\address[Second]{Autonomous Learning Robots (ALR), Institute for Anthropomatics and Robotics, Karlsruhe Institute of Technology, Germany, 
    \{%
        firstname%
    \}%
    .%
    \{%
      lastname%
    \}
    {@kit.edu}
}

\input{sections/00_abstract.tex}


\end{frontmatter}

\renewcommand{\vec}[1]{\ensuremath{{\underline{#1}}}}
\renewcommand{\rvec}[1]{\ensuremath{{\boldsymbol{\underline{#1}}}}}

\input{sections/01_intro.tex}
\input{sections/02_background.tex}
\input{sections/03_related_work.tex}
\input{sections/04_method.tex}
\input{sections/05_evaluation.tex}
\input{sections/06_conclusion.tex}


\balance
\bibliography{bib/literature.bib}             
                                                   







\end{document}

%% file: settings/packages.tex
\usepackage{etoolbox}

\usepackage{comment}

\usepackage{acronym}

\makeatletter
\def\input@path{{settings}} 
\usepackage{settings/isasmathmacros} 
\makeatletter

\usepackage{nicefrac}
\usepackage{array}
\usepackage{booktabs}

\usepackage[linesnumbered,ruled,vlined]{algorithm2e}
\AtBeginEnvironment{algorithm}{\DontPrintSemicolon}

\usepackage[x11names, svgnames, rgb]{xcolor} 
\usepackage{graphics} 
\usepackage[]{graphicx}  
\graphicspath{ {./images/**} } 

\usepackage[skip=2pt, ]{subcaption}

\usepackage{balance} 

\usepackage{stfloats} 

\usepackage{cleveref}

\crefname{figure}{Fig.}{Figs.} 
\crefname{section}{Sec.}{Secs.} 
\crefname{table}{Tab.}{Tabs.} 
\crefname{algorithm}{Alg.}{Algs.} 
\crefname{equation}{}{} 

\expandafter\def\csname ver@etex.sty\endcsname{3000/12/31}

\usepackage{autonum}

\usepackage{siunitx}
\usepackage{microtype}
\usepackage[all]{nowidow}
\usepackage[subtle,bibbreaks,bibnotes,mathspacing=normal]{savetrees}

%% file: settings/definitions.tex
\usepackage{bbold} 

\newcommand{\ie}{i.e.}

\newcommand{\eg}{e.g.}
\newcommand{\Eg}{E.g.}

\newcommand{\abs}[1]{{\ensuremath{\left \lvert #1 \right \rvert}}}

\newcommand{\indFunc}{\ensuremath{\mathds{1}}}

\definecolor{myblue}{named}{RoyalBlue}

\definecolor{mygreen}{named}{LimeGreen}

\newcommand{\cemPropPdf}{\ensuremath{{f}}} 

\newcommand{\dimControl}{\ensuremath{d_u}} 
\newcommand{\horizon}{\ensuremath{N}_{\mathrm{H}}} 

\newcommand{\numCEMIter}{\ensuremath{j}_\mathrm{max}} 

\newcommand{\optVarCemScalar}{\ensuremath{\xi}} 
\newcommand{\vVarCem}{\ensuremath{\vec{\optVarCemScalar}}} 
\newcommand{\rvVarCem}{\ensuremath{\rvec{\optVarCemScalar}}} 
\newcommand{\dimVarCem}{\ensuremath{d_{{\optVarCemScalar}}}} 
\newcommand{\evVarCem}{\ensuremath{\mean{\vec{\optVarCemScalar}}}} 
\newcommand{\paramCEM}{\ensuremath{\vtheta}} 
\newcommand{\weightFuncCEM}{\ensuremath{\varphi}} 

\newcommand{\lcdVarScalar}{\ensuremath{\xi}} 
\newcommand{\vVarLcd}{\ensuremath{\vec{\lcdVarScalar}}} 

\newcommand{\bufferSize}{\ensuremath{N_{\mathrm{Buffer}}}} 

%% file: settings/abbreviations.tex
\acrodef{AL}{active learning}
\acrodef{ANEES}{averaged normalized estimation error squared}
\acrodef{ANLL}{average negative log-likelihood}
\acrodef{BNN}{Bayesian neural network}
\acrodef{CDF}{cumulative density function}
\acrodef{CEM}{cross-entropy method}
\acrodef{CMA-ES}{covariance matrix adaptation evolution strategy}
\acrodef{CvM}{Cram\'er--von Mises}
\acrodef{DDP}{differential dynamic programming}
\acrodef{DM}{Dirac mixture}
\acrodef{DMD}{Dirac mixture density}
\acrodef{dsCEM}{deterministic sampling CEM}
\acrodef{dsMPPI}{deterministic sampling MPPI}
\acrodef{ECE}{expected calibration error}
\acrodef{EKF}{Extended Kalman Filter}
\acrodef{EM}{expectation--maximization}
\acrodef{ENCE}{expected normalized calibration error}
\acrodef{EP}{Expectation Propagation}
\acrodef{GM}{Gaussian mixture}
\acrodef{GEVR}{generalized error variance ratio}
\acrodef{GUCE}{generalized UCE}
\acrodef{iLQR}{iterative linear quadratic regulator}
\acrodef{iCEM}{improved CEM}
\acrodef{kdtree}{$k$-dimensional tree}
\acrodef{KBNN}{Kalman Bayesian Neural Networks}
\acrodef{KL}{Kullback--Leibler}
\acrodef{KS}{Kolmogorov--Smirnov}
\acrodef{LCD}{localized cumulative distribution}
\acrodef{LRKF}{Linear Regression Kalman filter}
\acrodef{MAP}{maximum a posteriori}
\acrodef{MC}{Monte Carlo}
\acrodef{MCMC}{Markov Chain Monte Carlo}
\acrodef{ML}{maximum likelihood}
\acrodef{MLE}{maximum likelihood estimator}
\acrodef{MPC}{model predictive control}
\acrodef{MPPI}{model predictive path integral}
\acrodef{MPPI Iterative}{MPPI Iterative}
\acrodef{MSE}{mean squared error}
\acrodef{MMSE}{minimum mean square error}
\acrodef{MNR}{matrix norm ratio}
\acrodef{MNRE}{matrix norm relative error}
\acrodef{MSER}{mean squared error ratio}
\acrodef{MV}{mean variance}
\acrodef{NCI}{noncredibility index}
\acrodef{NEES}{normalized estimation error squared}
\acrodef{NLL}{negative log-likelihood}
\acrodef{NGUCE}{normalized GUCE}
\acrodef{NUTS}{No-U-Turn Sampler}
\acrodef{ODE}{ordinary differential equation}
\acrodef{OCP}{optimal control problem}
\acrodef{PBP}{probabilistic backpropagation}
\acrodef{PCD}{projected cumulative distribution}
\acrodef{PDF}{probability density function}
\acrodef{QCE}{quantile calibration error}
\acrodef{RMSE}{root MSE}
\acrodef{RMV}{root MV}
\acrodef{RNN}{recurrent neural network}
\acrodef{S2KF}{Smart Sampling Kalman Filter}
\acrodef{SVI}{Stochastic Variational Inference}
\acrodef{TAGI}{Tractable Approximate Gaussian Inference}
\acrodef{UCE}{uncertainty calibration error}
\acrodef{UKF}{Unscented Kalman Filter}
\acrodef{VI}{Variational Inference}

%% file: sections/00_abstract.tex
\begin{abstract}                
    Sampling-based \ac{MPC} is effective for nonlinear systems but often produces non-smooth control inputs due to random sampling.
    To address this issue, we extend the \ac{MPPI} framework with deterministic sampling and improvements from \ac{CEM}--\ac{MPC}, such as iterative optimization, proposing \ac{dsMPPI}.
    This combination leverages the exponential weighting of \ac{MPPI} alongside the efficiency of deterministic samples.
    Experiments demonstrate that \ac{dsMPPI} achieves smoother trajectories compared to state-of-the-art methods.
    %
\end{abstract}

%% file: sections/01_intro.tex
\section{Introduction}
\label{sec:intro}

Sampling-based \ac{MPC} methods have gained significant attention in recent years due to their ability to handle complex nonlinear systems and nonconvex cost functions.
Similar to classical \ac{MPC}, sampling-based \ac{MPC} solves a finite-horizon \ac{OCP} at each time step, and applies the first control input of the optimized control sequence to the system.
Rather than relying on gradient-based optimization on the cost function, sampling-based \ac{MPC} models the control input sequence as a parameterized discrete-time stochastic process and employs sampling-based optimization methods to iteratively improve the parameters.
Specifically, samples of control input sequences are drawn from a proposal distribution, evaluated using the system dynamics and cost function, and then used to update the proposal distribution parameters. 
Using modern hardware, these steps can be performed efficiently in parallel.
Popular sampling-based \ac{MPC} methods include \ac{CEM}--\ac{MPC}~\citep{chuaDeepReinforcementLearning2018, pinneriSampleefficientCrossentropyMethod2021} and \ac{MPPI}~\citep{williamsInformationtheoretic2018, bhardwajSTORMIntegratedFramework2022}.

\begin{figure}
    \centering%
    \includegraphics[trim={0.425cm 0.078cm 0cm 0.23cm},clip, width=0.85\linewidth]{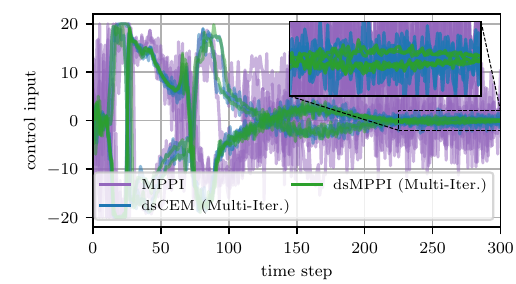}%
    \caption{Control input trajectories (five randomly initialized runs per method) for the cart-pole swing-up. 
    The proposed dsMPPI yields the smoothest inputs.
    }%
    \label{fig:eyecatcher:cart_pole:action_time_series}%
\end{figure}

A key challenge of sampling-based \ac{MPC} methods is that they typically yield non-smooth control inputs (see, \eg,~\cref{fig:eyecatcher:cart_pole:action_time_series} for \ac{MPPI}), 
as the proposal parameter update is approximated using a finite number of \emph{random} samples, leading to noisy estimates that worsen for small sample counts.
In real-world applications, this may cause problems, such as excessive actuator wear.
\Eg, noisy control inputs may cause the torque of a motor to frequently switch between large positive and large negative values within a short time span, leading to rapid velocity reversals.
Therefore, \emph{smooth control inputs are desirable}.
A common approach to addressing this issue is to apply low-pass filtering, 
to the optimized control inputs~\citep{williamsInformationtheoretic2018}.
However, this post-processing step leads to suboptimal performance, as it is not considered in the optimization.

In previous work, we proposed the \acfi{dsCEM} \citep{arXiv25_Walker}, which substitutes random sampling with deterministic samples generated by minimizing the modified \ac{CvM} distance~\citep{CDC09_HanebeckHuber}.
This yields samples with low discrepancy, which cover the search space without clustering or leaving huge gaps, as random sampling does.
This approach has been shown to significantly improve the smoothness of control inputs compared with standard \ac{CEM}.
However, \ac{dsCEM} still relies on \ac{CEM}'s elite set update rule, which performs hard selection, weighing the $N$ best samples equally, and disregarding all others.
In contrast, \ac{MPPI} uses a soft, exponential weighting scheme based on sample costs, enabling smoother updates.
%
In this paper, we propose combining the benefits of deterministic sampling with the \ac{MPPI}-style exponential weighting scheme.
We term this approach \acfi{dsMPPI} and show that it further improves the smoothness of control inputs compared to both \ac{MPPI} and \ac{dsCEM}, as illustrated in~\cref{fig:eyecatcher:cart_pole:action_time_series}.

\paragraph*{Contribution and Outline}%
First, in \cref{sec:background}, we provide a concise background on sampling-based \ac{MPC}  and review algorithmic improvements in \cref{sec:related_work}.
Then, in \cref{sec:proposed_method}, we propose \ac{dsMPPI}, a novel sampling-based \ac{MPC} algorithm that utilizes state-of-the-art improvements from both \ac{CEM} and \ac{MPPI} literature, combined with deterministic sampling.
Finally, in \cref{sec:evaluation}, we evaluate the performance of the proposed algorithm through extensive simulations.

\paragraph*{Notation}%
%
Vectors are denoted by underlined letters, \eg, $\vx$, random variables by boldface letters, such as $\rvx$, and matrices by boldface capitals, such as $\mA$.
The mean of a random variable is denoted by $\hat{\cdot}$, \eg, $\hat{\vx}$, covariance matrices by $\mC$, and diagonal matrices by $\diag(\cdot)$.
The indicator function $\indFunc_{A}$ is \num{1} if $A$ is true, and \num{0} otherwise.

%% file: sections/02_background.tex
\section{Background}
\label{sec:background}

We consider a discrete-time dynamical system given by
\begin{align}
    \label{eq:system_dynamics}
    \vx_{k+1} & = \va_k(\vx_k, \vu_k) \enspace,
\end{align}
and a finite-horizon \ac{OCP} with cumulative cost
\begin{align}
    \label{eq:ocp_cost}
    J_k & = g_{\horizon}(\vx_{k + \horizon}) + \sum_{n=k}^{k+\horizon-1} g_n(\vx_n, \vu_n)  \enspace,
\end{align}
where $k$ is the current time step, $\horizon$ is the prediction horizon, $\vx_k \in \IR^{d_x}$ is the system state, $\vu_n \in \IR^{d_u}$  is the control input, $g_n(\vx_n, \vu_n)\colon \IR^{d_x} \times \IR^{d_u} \to \IR$ is the stage cost at time step $n$, and $g_{\horizon}(x_{k+\horizon})\colon \IR^{d_x} \to \IR$ is the terminal cost. 
Note that no further assumptions are made about the dynamics or cost functions, which can both be time-variant.
The goal of the \ac{OCP} is to find the optimal control sequence $\vu_{k:k+\horizon-1}^* = \vu_k^*, \vu_{k+1}^*, \ldots, \vu_{k+\horizon-1}^*$ that minimizes the cumulative cost $J_k$ subject to the system dynamics in \eqref{eq:system_dynamics}.
For brevity, we denote the flattened control sequence $\vu_{k:k+\horizon-1}$ as $\vVarCem \in \IR^{\dimVarCem}$ in the following, \ie, $\vVarCem = [\vu_k\T, \vu_{k+1}\T, \ldots, \vu_{k+\horizon-1}\T]\T \in \IR^{\horizon \cdot d_u}$, and the corresponding cumulative cost $J_k$ as $J(\vVarCem)$.

In sampling-based \ac{MPC} methods, control sequences are drawn from a proposal distribution $\cemPropPdf(\vVarCem; \vtheta_j)$, parameterized by $\vtheta_j$, where $j \in \{0, \ldots, \numCEMIter-1\}$ denotes the iteration index and $\numCEMIter$ is the number of iterations.
The proposal is \emph{iteratively} updated to generate control sequences with low costs that are near the optimal solution to the \ac{OCP}.
\Eg, in the \ac{CEM} proposed by~\cite{rubinsteinCrossEntropy2004}, the proposal distribution is iteratively updated by solving
\begin{align}
        \paramCEM_{j+1} &= \argmin_{\vtheta'} D_{\mathrm{KL}}\mleft(\cemPropPdf_j^*(\vVarCem) \| \cemPropPdf(\vVarCem;\vtheta') \mright) 
        \\
        & = \argmin_{\vtheta'} H\mleft(\cemPropPdf_j^*(\vVarCem), \cemPropPdf(\vVarCem;\vtheta') \mright) \enspace, 
\end{align}
where $D_{\mathrm{KL}}(\cdot \| \cdot)$ is the \ac{KL} divergence, and $H(\cdot, \cdot)$ 
denotes the cross-entropy. 
The parameters are updated by minimizing the \ac{KL} divergence between an optimal~\emph{importance sampling} distribution $\cemPropPdf_j^*$ for iteration $j$ and the proposal distribution $\cemPropPdf(\cdot;\vtheta')$.
This optimal importance sampling distribution $\cemPropPdf_j^*(\vVarCem)$, which minimizes the variance of the importance sampling estimator~\citep{rubinsteinCrossEntropy2004}, is given by
$
    \cemPropPdf_j^*(\vVarCem) \propto \weightFuncCEM \left(J(\vVarCem) \right) \, \cemPropPdf(\vVarCem;\paramCEM_j) 
$,
where the weighting function $\varphi(J(\vVarCem))$ assigns higher probability to control sequences with lower costs.
Rewriting the cross-entropy term as an expectation w.r.t. the proposal distribution $\cemPropPdf(\vVarCem;\paramCEM_j)$, expresses the update rule as
\begin{align}
    \label{eq:cem_update_expectation}
    \paramCEM_{j+1} = \argmin_{\vtheta'} - \Eop_{\cemPropPdf(\vVarCem; \paramCEM_j)} \big\{ \weightFuncCEM \left(J(\vVarCem) \right) \log \cemPropPdf(\vVarCem; \vtheta') \big\} \enspace.
\end{align}
The expectation in \cref{eq:cem_update_expectation} is approximated using samples $\{\vVarCem^{(i)}\}_{i=1}^{N}$ drawn from $\cemPropPdf(\vVarCem; \paramCEM_j)$, and solving for $\paramCEM_{j+1}$ adapts the distribution to generate lower-cost samples in the next iteration.
%
%
For Gaussian proposal distributions, where $\paramCEM$ consists of the mean $\evVarCem$ and covariance matrix $\mC$, the update \cref{eq:cem_update_expectation} admits a closed-form solution given by the weighted sample moments
\begin{align}
    \label{eq:sample_mean}
    \evVarCem_{j+1} & = \sum_{i=1}^{N} \frac{w^{(i)}}{\eta} \vVarCem^{(i)}  \enspace,\\
    \label{eq:sample_cov}
    \mC_{j+1} & = \sum_{i=1}^{N} \frac{w^{(i)}}{\eta} (\vVarCem^{(i)} - \evVarCem_{j+1})(\vVarCem^{(i)} - \evVarCem_{j+1})\T \enspace,
\end{align}
where $w^{(i)} = \weightFuncCEM (J(\vVarCem^{(i)}))$ and $\eta = \sum_{i=1}^{N} w^{(i)}$ are the weights and the normalization constant, respectively.

A common special case is the use of 
$\weightFuncCEM(\cdot) = \indFunc_{J(\vVarCem)\le \gamma_j}$, that is, selecting
an elite set of samples whose costs fall below a certain threshold $\gamma_j$. 
The update rule then reduces to sample moments computed only over the elite set, where each elite sample is weighted equally.

Using an exponential weighting $\varphi(J(\vVarCem)) = \exp(\nicefrac{-J(\vVarCem)}{\lambda})$ with inverse temperature $\lambda > 0$ relates the \ac{CEM} to \ac{MPPI}~\citep{williamsInformationtheoretic2018}.
Standard \ac{MPPI} computes the control input $\vu_k$ directly via weighted averaging according to \cref{eq:sample_mean} within only one iteration, \ie, without iteratively updating the proposal distribution.
While \ac{MPPI} connects to optimal control and information-theoretic principles~\citep{williamsInformationtheoretic2018}, it imposes restrictive assumptions such as quadratic control costs.
In contrast, the \ac{CEM} formulation with exponential weights is more flexible and allows for arbitrary cost functions.

%% file: sections/03_related_work.tex
\section{Related Work}
\label{sec:related_work}

We now review relevant literature on sampling-based \ac{MPC}.
In this context, \ac{CEM} typically denotes methods that employ an elite set, while \ac{MPPI} refers to methods using an exponential weighting scheme.
For the remainder of this paper, we adopt this convention. 

\subsection{Improvements to CEM--MPC}



Standard improvements include warmstarting the proposal distribution based on the previous solution~\citep{chuaDeepReinforcementLearning2018,pinneriSampleefficientCrossentropyMethod2021} and applying momentum smoothing to stabilize parameter updates~\citep{rubinsteinCrossEntropy2004,pinneriSampleefficientCrossentropyMethod2021}.
To address the high dimensionality of control sequences, the covariance matrix is commonly assumed to be diagonal, which simplifies estimation and sampling~\citep{hafnerLearning2019}. However, doing so neglects temporal correlations and sacrifices the smoothness of the control inputs.


In the \acfi{iCEM}, \citet{pinneriSampleefficientCrossentropyMethod2021} proposed a notable improvement that considers temporal correlations in the control sequence through random colored noise sampling with a power spectral density $\text{PSD}(f) \propto \nicefrac{1}{f^\beta}$, where $f$ is the frequency and $\beta$ controls the noise color.
This allows for generating smoother control sequences while maintaining computational efficiency by updating only the marginal variances instead of the full covariance matrix.
Additionally, \ac{iCEM} uses a buffer to preserve a fraction of the previous elite samples, enhancing sample efficiency by maintaining effective solutions across iterations.
Furthermore, the best solution found is returned after the \ac{MPC} step.

%


In our previous work~\citep{arXiv25_Walker}, we proposed the \ac{dsCEM}, which uses deterministic samples generated offline by minimizing the modified \ac{CvM} distance~\citep{CDC09_HanebeckHuber} between a multivariate standard normal distribution and its sample approximation.
These optimal samples are stored and transformed online to match the current proposal distribution.
Deterministic sampling improves performance and sample efficiency compared to random sampling. 
However, using the same sample set for every iteration would result in poor exploration.
To enhance exploration, we introduced variation schemes, such as selecting a different subset from a larger pool of precalculated samples for each iteration or applying random rotations to the stored samples.
In addition, to capture temporal structure without incurring the computational cost of a full covariance matrix, we proposed updating only the marginal variances in each \ac{CEM} iteration while using fixed time correlations.

\subsection{Improvements to \acs{MPPI}}
For numerical stability, \ac{MPPI} typically employs cost shifting relative to the minimum cost over all sampled trajectories \citep{williamsInformationtheoretic2018}.
While standard \ac{MPPI} performs a single weighted mean update ({$\numCEMIter=1$}) per \ac{MPC} time step, iterative versions have been proposed.
\Eg, \citep{bhardwajSTORMIntegratedFramework2022} integrated an \ac{MPPI}-like exponential weighting into an iterative optimization scheme similar to \ac{CEM}.
%
A critical hyperparameter in \ac{MPPI} is the inverse temperature $\lambda$, which controls the ``sharpness'' of the weights.
Typically, $\lambda$ is fixed or manually tuned.
Recently, \cite{pezzatoSamplingbasedModelPredictive2025} proposed a self-adaptation heuristic for $\lambda$ that adjusts the temperature in each time step based on the costs of the sampled trajectories.

%% file: sections/04_method.tex
\section{Proposed Method}
\label{sec:proposed_method}

We propose an improved sampling-based \ac{MPC} algorithm that combines the strengths of \ac{CEM} and \ac{MPPI} with deterministic sampling.
The key components of our proposed method, referred to as \ac{dsMPPI},  are detailed below, and summarized in~\cref{alg:ds_sample_mpc_alg}.

\paragraph*{Iterative Update}
Similar to \ac{CEM}--\ac{MPC}, we use multiple iterations to update a proposal distribution for the control sequence. 
In each iteration $j$, 
the importance-weighted average known from \ac{MPPI} is used,
including a cost shift for numerical stability.
That is, the weights are given by
\begin{align}
    \label{eq:imp_weights}
    w^{(i)} = \frac{1}{\eta} \exp\left(-\frac{1}{\lambda} \left(J\left(\vVarCem^{(i)}\right) - \rho \right) \right) \enspace,
\end{align}
where $\rho = \min \{ J(\vVarCem^{(i)}) \}_{i=1}^{N}$ is the minimum cost over all trajectories, and 
$\eta$
is the weight normalization constant.
As in \citep{hafnerLearning2019}, we assume a Gaussian proposal distribution for the control sequence, updating only the marginal variances to keep the dimensionality of the parameter space low.
The update for the marginal variances $(\vsigma^{\prime}_j)^{2}$ of $\mC_j^{\prime}$ then becomes
\begin{align}
    \label{eq:prop_sample_cov}
    (\vsigma^{\prime}_j)^{2} &=  \sum_{i=1}^{N} w^{(i)} (\vVarCem_j^{(i)} - \evVarCem_{j}^{\prime})^{2} \enspace,
\end{align}
where $(\cdot)^{2}$ denotes the element-wise square. 
Note that the quantities denoted by $\cdot^{\prime}$ are intermediate quantities used for momentum smoothing.

\paragraph*{Momentum Smoothing}
To stabilize the optimization and prevent premature convergence, we apply momentum smoothing to both the mean and covariance updates, as used in \citep{bhardwajSTORMIntegratedFramework2022}.
Both the mean and covariance are updated via
\begin{align}
    \label{eq:cem_momentum_mean}
    \paramCEM_{j+1} = \alpha \, \paramCEM_{j} + (1-\alpha) \, \paramCEM_{j}^{\prime} \enspace,
\end{align}
where $\paramCEM_{j}^{\prime} = (\evVarCem_{j}^{\prime}, \mC_{j}^{\prime})$ contains the weighted sample mean and covariance, and $\alpha \in [0,1)$ is a momentum factor.

\paragraph*{Adaptive Temperature}
To overcome the limitations of a fixed inverse temperature parameter $\lambda$, we use a heuristic adaptation rule.
Similar to~\cite{pezzatoSamplingbasedModelPredictive2025}, we use
\begin{align}
    \label{eq:temp_adaption}
    \lambda_{j+1} = \begin{cases} 
        0.9 \, \lambda_j & \text{if~} \eta > \eta_{\mathrm{max}}
        \\
        1.2 \, \lambda_j & \text{if~} \eta < \eta_{\mathrm{min}}
        \\
        \lambda_j & \text{otherwise}        
        \end{cases} \enspace,
\end{align}
where $\eta$ is the normalization constant from \eqref{eq:imp_weights},
and $\lambda_j$ and $\lambda_{j+1}$ are the inverse temperatures in iterations $j$ and $j+1$, respectively.
Thresholds $\eta_{\mathrm{min}}$ and $\eta_{\mathrm{max}}$ are chosen empirically to ensure smooth behavior, and \citet{pezzatoSamplingbasedModelPredictive2025} found that the range $[5, 10]$ performs well.

\begin{figure}[t]
    \centering
    \begin{subfigure}{0.45\columnwidth}
        \centering
        \includegraphics[width=0.75\columnwidth]{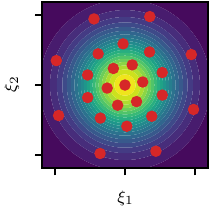}%
        \caption{Standard normal samples}%
        \label{fig:lcd_samples:standard}%
    \end{subfigure}
    \begin{subfigure}{0.45\columnwidth}
        \centering
        \includegraphics[width=0.75\columnwidth]{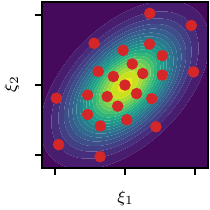}%
        \caption{Transformed samples}%
        \label{fig:lcd_samples:transformed}%
    \end{subfigure}%
    \caption{Example of two-dimensional deterministic samples.}
    \label{fig:lcd_samples}%
\end{figure}

\paragraph*{Deterministic Sampling with Variation}
Building upon our previous work~\citep{arXiv25_Walker}, we incorporate deterministic samples to enhance the smoothness of the control inputs and improve sample efficiency.
In each iteration, precomputed deterministic samples \citep{CDC09_HanebeckHuber} $\{ {\vVarLcd}^{(i)}_{\mathrm{SN}} \}$ from a isotropic standard normal $\rvVarCem \sim \mathcal{N}(\vVarCem; \vzero, \mI)$ are transformed 
to match the current proposal distribution parameters $\paramCEM_j = (\evVarCem_j, \mC_j)$.
The transformation for each sample $\vVarCem_j^{(i)}$ is given by%
\begin{align}
    \label{eq:det_sample_transform}
    \vVarCem_j^{(i)} & = \evVarCem_j + \mL_j \, {\vVarLcd}^{(i)}_{\mathrm{SN}} \enspace,
\end{align}
where $\mL_j$ is the square root of $\mC_j$ such that $\mC_j = \mL_j \mL_j\T$.
\cref{fig:lcd_samples} provides a visual intuition of this concept.

Since reusing the exact same deterministic samples in every iteration can lead to poor exploration, we implement two variation schemes to enhance diversity.
A simple method is generating deterministic samples from an isotropic Gaussian with dimension $\dimVarCem \cdot \numCEMIter$ and selecting different subsets of size $\dimVarCem$ for each iteration $j$,  as proposed in \citep{arXiv25_Walker}. 
We refer to this approach as the \emph{multi-iteration} method.
Since storing a large pool of deterministic samples can be memory-intensive, we propose a second variation scheme, referred to as the \emph{permutation} method, that randomly permutes the dimensions of each deterministic sample in every iteration.
\Eg, in one iteration the dimension of the samples is interpreted as $[u_1, u_2, \ldots, u_{\dimControl \cdot \horizon}]$, whereas in the next iteration the dimensions are randomly permuted to, \eg, $[u_3, u_1, u_n, \ldots]$.
This approach requires storing only a single set of deterministic samples while still promoting exploration through dimension shuffling (this only requires a discrete random number generator).
Furthermore, the standard normal samples remain optimal with respect to the modified \ac{CvM} distance~\citep{CDC09_HanebeckHuber} since an isotropic Gaussian is permutation-invariant.

\paragraph*{Time Correlations}
For smooth control inputs, it is beneficial to introduce correlations between control inputs at different time steps.
Following \cite{arXiv25_Walker}, the covariance matrix is constructed as $\mC_j = \diag(\vsigma_j) \mC_{\rho} \diag(\vsigma_j)$, where $\mC_{\rho}$ is the fixed time-correlation matrix and $\vsigma_j$ is the vector of marginal standard deviations. 
The transformation in \cref{eq:det_sample_transform} then uses 
$
    \mL_j = \diag(\vsigma_j) \mA_{\rho} 
$,
where $\mA_{\rho}$ is the matrix square root of $\mC_{\rho}$.
This approach allows for capturing temporal correlations while keeping the number of parameters manageable by only updating the marginal variances $(\vsigma_j)^2$ at each iteration.
Since the time-correlation structure is fixed, we calculate $\mA_{\rho}$ offline to reduce computational overhead.
As in \cite{arXiv25_Walker}, we use a correlation structure derived from colored noise, following the power spectral density law, which has been shown to produce smooth control trajectories suitable for many robotic applications~\citep{eberhard2023pink}.
The initial time correlation matrix $\mC_{\mathrm{t}}$ is derived from the power spectral density of colored noise 
using the Wiener--Khinchin theorem~\cite[pp. 576--578]{kay1993fundamentals}.

For multivariate control inputs, the overall correlation matrix $\mC_{\rho}$ is constructed by
%
$
    \mC_{\rho} = \mC_{\mathrm{t}} \otimes \mC_{\mathrm{sp}} 
$,
where $\otimes$ denotes the Kronecker product, 
and $\mC_{\mathrm{sp}}$ captures spatial correlations between different control input dimensions at the same time step. 
For simplicity, in this work we assume independent control input dimensions and set $\mC_{\mathrm{sp}} = \mI$.




\paragraph*{Further improvements}
To ensure that the sampled control inputs respect input bound constraints, clamping is applied. 
This can be seen as modified nonlinear system dynamics~\citep{williamsInformationtheoretic2018}
\begin{align}
    \label{eq:clamped_dynamics}
    \vx_{k+1} = \va_k \left(\vx_k, \operatorname{clamp}\left(\vu_k, \vu_{\min}, \vu_{\max} \right) \right) \enspace,
\end{align}
where $\operatorfont{clamp}(\cdot)$ restricts the control inputs to the specified bounds $\vu_{\min}$ and $\vu_{\max}$. 

Additionally, we use a buffer of size $\bufferSize$ to keep the best trajectories from the previous iteration, as in \citet{pinneriSampleefficientCrossentropyMethod2021}.
This improves sample efficiency and allows returning the best found control input instead of the proposal mean at the end of the \ac{MPC} step.
Furthermore, as in \citet{pinneriSampleefficientCrossentropyMethod2021}, we perform a warm start of the proposal distribution at each time step by shifting the mean and buffered trajectories based on the previous solution. 
Shifted sequences are given by
\begin{align}
    \label{eq:shift_warmstart}
    \vVarCem = [\vu_{k-1, 1:\horizon-1}\T, \vu_{k-1, \horizon-1}\T]\T \enspace,
\end{align}
where $\vu_{k-1, n}$ is the control input at stage $n$ within the horizon from the previous \ac{MPC} step ($k-1$), and the last control input is repeated to maintain the horizon length.


\IncMargin{1.5em} 
\begin{algorithm}[t]
    \small
    \caption{dsMPPI Step}
    \label{alg:ds_sample_mpc_alg} 
    \SetKwFunction{FMain}{CEM-MPC-Step}
    \SetKwInOut{Input}{Input}
    \Input{%
        State $\vx_k$, initial parameters $\paramCEM_0 = (\evVarCem_0, \mC_0)$, buffered samples from previous step
    }%
    \SetKwInOut{Output}{Optimal control input $\vu_k^*$}
    Warm start last proposal mean and buffered samples \cref{eq:shift_warmstart}\;
    \For{$j \gets 0$ \KwTo $\numCEMIter-1$}{
        Transform deterministic samples \cref{eq:det_sample_transform}\; 
        Add saved samples from buffer\;
        Trajectory shooting using \cref{eq:clamped_dynamics}~\tcp{\small{parallel}}
        Evaluate costs 
        using~\cref{eq:ocp_cost}~\tcp{\small{parallel}}
        Calculate weights $w^{(i)}$ using \cref{eq:imp_weights}\;
        Calculate weighted sample moments 
        \cref{eq:prop_sample_cov} 
        \;
        Update proposal distribution using \cref{eq:cem_momentum_mean}\;
        Update temperature parameter $\lambda_j$ using \cref{eq:temp_adaption}\;
        Add $\bufferSize$ best samples to buffer\;
    }
    \Return first control $\vu_k^*$ from the best sequence
\end{algorithm}
\DecMargin{1.5em}

%% file: sections/05_evaluation.tex
\section{Experiments}
\label{sec:evaluation}


%

In this section, we evaluate the performance of the proposed \ac{dsMPPI} algorithm and the novel permutation variation scheme.
We compare these contributions against our previous work, \ac{dsCEM} \citep{arXiv25_Walker}, and standard \ac{MPPI} \citep{williamsInformationtheoretic2018}.
Additionally, we perform an ablation study using a random sampling variant of our proposed \ac{dsMPPI} algorithm, denoted as \emph{\acs{MPPI Iterative}\acused{MPPI Iterative}}, to isolate the benefits of deterministic sampling.
We conduct experiments on the cart-pole swing-up and the truck backer-upper tasks.
For both tasks, quadratic stage cost functions of the form $g_n(\vx_n, u_n) = (\vx_n - \vx_{\mathrm{g}})\T \mQ (\vx_n - \vx_{\mathrm{g}}) + \vu_n\T \mR \vu$ are used, where $\vx_{\mathrm{g}}$ is the goal state, and $\mQ$ and $\mR$ are state and control weights, respectively. 
The terminal cost function is set to $g_{\horizon}(\vx_{k+\horizon}) = (\vx_{k+\horizon} - \vx_{\mathrm{g}})\T \mQ_{\horizon} (\vx_{k+\horizon} - \vx_{\mathrm{g}})$.
The task-specific parameters are summarized in \cref{tab:task_parameters}.

\begin{table}[t]
    \centering
    \caption{Task-specific parameters.}
    \label{tab:task_parameters}
    \setlength{\tabcolsep}{0.5pt}
    {
    \small
    \begin{tabular}{lcc}
        \toprule
        Parameter                   & Cart-Pole Swing-Up                                & Truck Backer-Upper \\
        \midrule
        Control limits              & $u \in [-20, 20]$                   & $u_1, u_2 \in [-1, 1]$ \\
        Goal state $\vx_{\mathrm{g}}$ ($\tilde{\vx}_{\mathrm{g}}$)  & $[0, 0, 1, 0, 0]\T$ &  $\vzero$           \\
        State weights $\mQ$         & $\diag([0.1, 0.1, 1, 0.1, 0.1])$                  &  $\diag([0.01, 0.5, 5, 0.01])$     \\
        State weights $\mQ_{\horizon}$ & $\diag([10, 0.1, 10, 0.1, 0.1])$            &   $\diag([0.1, 1, 10, 0.1])$            \\
        Control weights $\mR$        & \num{1e-4}                                        & $\diag([\num{1e-3}, \num{5}])$ \\
        Noise color $\beta$         & \num{0.25}                                           & \num{1}       \\
        Initial $\evVarCem_{0}$     & $\vzero$                                          & $\vzero$      \\
        Initial $\vsigma_0$         & $\num{10}\cdot\vones$                             & $\vones$      \\
        Iterations $\numCEMIter$    & \num{3}                                           & \num{3}       \\
        Horizon $\horizon$          & \num{30}                                        & \num{15}      \\
        Buffer size $\bufferSize$   & \num{3}                                           & \num{3}       \\
        Total time steps $T$        & \num{300}                                         & \num{100}     \\
        \bottomrule
    \end{tabular}
    }
\end{table}

We evaluate controller performance using cumulative cost and control input smoothness.
The cumulative cost is the sum of stage costs $g_k(\vx_k, \vu_k)$ over the entire simulation.
As in \citep{arXiv25_Walker}, the smoothness is measured using
    $\sum_{k=1}^{T-1} \| \vu_{k} - \vu_{k-1} \|^2 $,
where a lower value indicates a smoother trajectory. 
Since smoothness near the goal state is particularly important to avoid oscillations around the set point, we also report the smoothness measure restricted to the second half of the trajectory (time steps $\lceil\nicefrac{T}{2}\rceil$ to $T-1$), denoted as \emph{settled smoothness}.
Each experiment uses sample sizes ranging from \num{20} to \num{300}, and each experiment is repeated \num{100} times with different random seeds to randomly sample the initial states. The median and interquartile range per setting are reported.

\paragraph*{Cart-Pole Swing-Up}
\label{sec:cart_pole}
The task involves swinging a pendulum mounted on a cart from a downward to an upright position while balancing it there. We use the same experimental setup as~\citet{arXiv25_Walker}, where the state is $\vx = [\mathsf{x}, \dot{\mathsf{x}}, \phi, \dot{\phi}]\T$, with cart position $\mathsf{x}$, pole angle $\phi$,
and their respective (angular) velocities. 
The costs are evaluated using an augmented state vector $\tilde{\vx} = [x_1, x_2, \cos(x_3), \sin(x_3), x_4]$ to account for the periodicity of the angle $x_3$.
%
%
%
Results for the cart-pole swing-up are shown in \cref{fig:cart_pole_results}.
Control input trajectories for \num{300} samples are visualized in \cref{fig:eyecatcher:cart_pole:action_time_series}.

\newcommand{\ifacImageScaling}{0.9}
\begin{figure}%
    \centering%
    \newcommand{\vspaceBelowSubfigure}{\vspace{4pt}}
    \begin{subfigure}{\columnwidth}
        \centering%
        \includegraphics[trim={0.175cm 0.078cm 0cm 0.15cm},clip,width=\ifacImageScaling\linewidth]{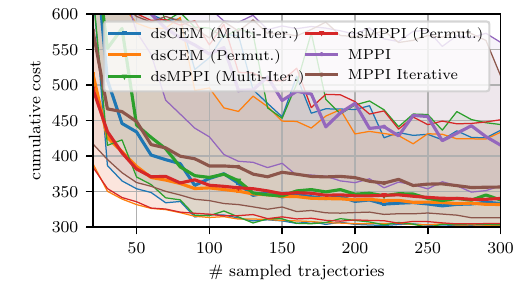}%
        \caption{Cumulative costs (lower is better)}%
        \label{fig:cart_pole:cum_cost}%
    \end{subfigure}
    \begin{subfigure}{\columnwidth}
        \centering%
        \includegraphics[trim={0.175cm 0.078cm 0cm 0.15cm},clip,width=\ifacImageScaling\linewidth]{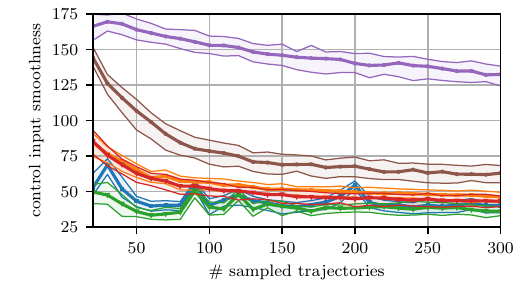}%
        \caption{Smoothness (lower is better)}%
        \label{fig:cart_pole:action_smoothness}%
    \end{subfigure}
    \begin{subfigure}{\columnwidth}
        \centering%
        \includegraphics[trim={0.175cm 0.078cm 0cm 0.15cm},clip,width=\ifacImageScaling\linewidth]{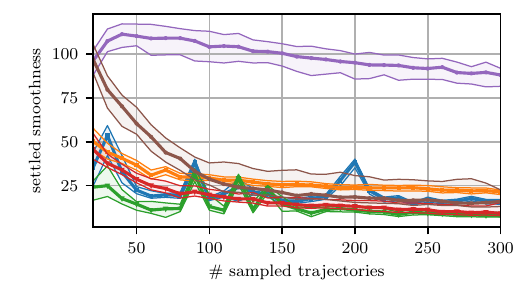}%
        \caption{Settled smoothness (lower is better)}%
        \label{fig:cart_pole:action_smoothness_settled}%
    \end{subfigure}
    \caption{Results for the cart-pole swing-up.} 
    \label{fig:cart_pole_results}%
\end{figure}%
\begin{figure}%
    \centering%
    \newcommand{\vspaceBelowSubfigure}{\vspace{4pt}}
    \begin{subfigure}{\columnwidth}
        \centering%
        \includegraphics[trim={0.175cm 0.078cm 0cm 0.15cm},clip,width=\ifacImageScaling\linewidth]{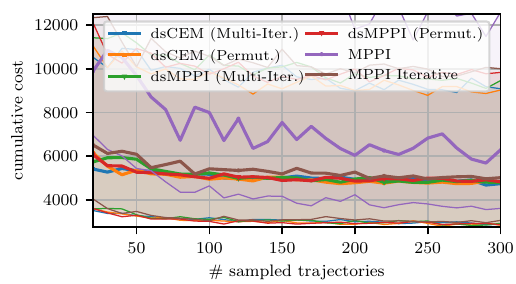}%
        \caption{Cumulative costs (lower is better)}%
        \label{fig:truck:cum_cost}%
    \end{subfigure}
    \begin{subfigure}{\columnwidth}
        \centering%
        \includegraphics[trim={0.175cm 0.078cm 0cm 0.15cm},clip,width=\ifacImageScaling\linewidth]{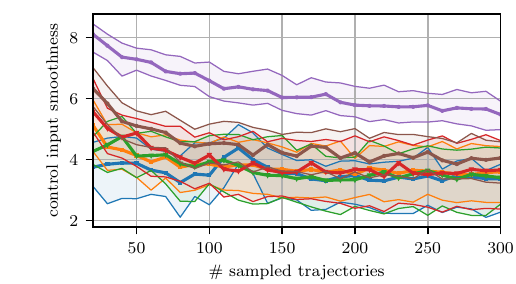}%
        \caption{Smoothness (lower is better)}%
        \label{fig:truck:action_smoothness}%
    \end{subfigure}
    \begin{subfigure}{\columnwidth}
        \centering%
        \includegraphics[trim={0.175cm 0.078cm 0cm 0.15cm},clip,width=\ifacImageScaling\linewidth]{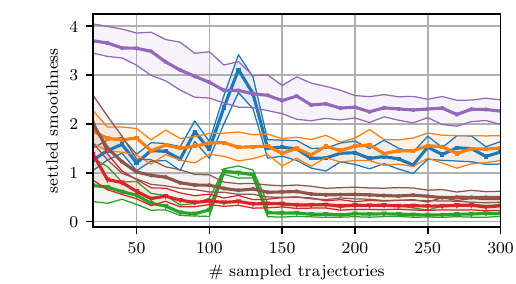}%
        \caption{Settled smoothness (lower is better)}%
        \label{fig:truck:action_smoothness_settled}%
    \end{subfigure}
    \caption{Results for the truck backer-upper.} 
    \label{fig:truck_results}%
\end{figure}%

\paragraph*{Truck Backer-Upper}
\label{sec:truck_backer_upper}
For the truck backer-upper \citep{schoenauerNeurogeneticTruckBackerupper1994}, the discrete-time dynamics reads
\newcommand{\stateX}[1]{\mathsf{x}_{#1}}
\newcommand{\stateY}[1]{\mathsf{y}_{#1}}
\newcommand{\thetaS}[1]{\theta_{\mathrm{S}, #1}}
\newcommand{\thetaC}[1]{\theta_{\mathrm{C}, #1}}
\newcommand{\uSteer}[1]{\tilde{u}_{1, #1}}
\newcommand{\uVel}[1]{\tilde{u}_{2, #1}}
\newcommand{\uSteerUnscaled}[1]{{u}_{1, #1}}
\newcommand{\uVelUnscaled}[1]{{u}_{2, #1}}
\newcommand{\uSteerMax}{\SI{70}{\degree}}
\newcommand{\uVelMax}{\SI{3}{\meter}}
\newcommand{\lenS}{L_{\mathrm{S}}}
\newcommand{\lenC}{L_{\mathrm{C}}}
\begin{align}
    \label{eq:truck}
    \begin{bmatrix}
    \stateX{k+1} \\
    \stateY{k+1} \\
    \thetaS{k+1} \\
    \thetaC{k+1} \\
    \end{bmatrix}
    = \begin{bmatrix}
    \stateX{k} - B \cdot \cos(\thetaS{k})\\
    \stateY{k} - B \cdot \sin(\thetaS{k})\\
    \thetaS{k} - \arcsin \left(  \frac{A \cdot \sin(\thetaC{k} - \thetaS{k}) }{\lenS} \right) \\
    \thetaC{k} + \arcsin \left(  \frac{\uVel{k} \cdot \sin (\uSteer{k})}{\lenS+ \lenC}  \right) \\
    \end{bmatrix}
\end{align}
where $\stateX{k}, \stateY{k}$ are the rear axle coordinates in \si{\meter} and  $\thetaS{k}, \thetaC{k}$ are the trailer and cab angles, respectively, relative to the $\stateX{}$-axis.
The control inputs are the steering angle $\uSteer{k}$ and velocity $\uVel{k}$.
The trailer and truck lengths are $\lenS = \SI{14}{\meter}$ and $\lenC = \SI{6}{\meter}$, respectively, with $A = \uVel{k} \, \cos(\uSteer{k})$ and $B = A \, (\thetaC{k} - \thetaS{k})$.
A jackknife constraint clamps $\abs{\thetaC{k}}$ to $\SI{90}{\degree}$ if the difference $\abs{\thetaS{k} - \thetaC{k}}$ exceeds $\SI{90}{\degree}$.
The optimization variables are the normalized inputs $\uSteerUnscaled{k}, \uVelUnscaled{k} \in [-1, 1]$, scaled by the maximums $\uSteerMax$ and $\uVelMax$ per time step, respectively.
Initial states are sampled uniformly from $\stateX{0} \in [80, 100]\,\si{\meter}$, $\stateY{0} \in [-50, 50]\,\si{\meter}$, and $\thetaS{0}, \thetaC{0} \in [-90, 90]\si{\degree}$.
The task is to back the trailer into the origin $(0,0)$ with both angles at $0\si{\degree}$, \ie, $\vx_{\mathrm{g}} = \vzero$.
Results for the truck backer-upper are shown in \cref{fig:truck_results}.


\paragraph*{Comparison of MPC Algorithms}
\label{sec:results_comparison}
We first compare the proposed \ac{dsMPPI} against \acl{MPPI Iterative}, \ac{dsCEM}, and standard \ac{MPPI}.
Standard \ac{MPPI} performs significantly worse than the iterative methods in both tasks.
Comparing the iterative methods, \ac{dsCEM} generally achieves slightly lower cumulative costs than \ac{dsMPPI} (\cref{fig:cart_pole:cum_cost}, \cref{fig:truck:cum_cost}).
However, \ac{dsMPPI} still yields slightly lower cumulative costs than its random sampling counterpart \ac{MPPI Iterative} in almost all settings, demonstrating the benefits of deterministic sampling.

In terms of control input smoothness, \ac{dsMPPI} demonstrates superior performance.
In the cart-pole swing-up, \ac{dsMPPI} consistently yields smoother controls for large sample sizes than both \ac{dsCEM} and \ac{MPPI Iterative} (\cref{fig:cart_pole:action_smoothness}).
This advantage is even more pronounced in the settled smoothness (\cref{fig:cart_pole:action_smoothness_settled}), where \ac{dsMPPI} outperforms all other methods.
This is visually confirmed by the control input trajectories in \cref{fig:eyecatcher:cart_pole:action_time_series}.
Similar results are obtained for the truck backer-upper (\cref{fig:truck:action_smoothness}, \cref{fig:truck:action_smoothness_settled}), where \ac{dsMPPI} consistently outperforms \ac{dsCEM}, \ac{MPPI Iterative}, and standard \ac{MPPI}.
This confirms that combining deterministic sampling with \ac{MPPI}'s exponential weighting effectively mitigates chattering while maintaining competitive costs.

\paragraph*{Comparison of Variation Schemes}
\label{sec:results_variation}
%
In terms of cumulative cost, the proposed permutation scheme slightly outperforms the multi-iteration scheme from \citep{arXiv25_Walker}. 
For the cart-pole swing-up (\cref{fig:cart_pole:cum_cost}), \ac{dsMPPI} and \ac{dsCEM} using the permutation scheme achieve lower costs than their multi-iteration counterparts, particularly for low sample sizes.
A similar trend is observed in the truck backer-upper (\cref{fig:truck:cum_cost}), where the permutation scheme yields competitive or lower costs.

Regarding control input smoothness, however, the multi-iteration scheme tends to produce smoother trajectories.
In the cart-pole swing-up (\cref{fig:cart_pole:action_smoothness}), the multi-iteration scheme consistently yields better smoothness for both \ac{dsMPPI} and \ac{dsCEM}.
This trade-off is also evident in the truck backer-upper (\cref{fig:truck:action_smoothness}), where the multi-iteration scheme achieves the best smoothness for \ac{dsMPPI}.
Overall, the permutation scheme offers a cost advantage, while the multi-iteration scheme favors smoothness.

\paragraph*{Computation Time Comparison}
\Cref{tab:computation_times} compares the computation times per control step, evaluated for \num{100} samples. 
Standard \ac{MPPI} is the fastest method due to its non-iterative nature.
Crucially, all iterative methods exhibit similar computation times, confirming that the proposed deterministic sampling scheme incurs no additional computational overhead. 

\begin{table}[t]
    \centering    
    \caption{Computation times per control step.}
    \label{tab:computation_times}
    \setlength{\tabcolsep}{2pt}
    {
    \small
    \sisetup{round-mode=places,round-precision=1, scientific-notation=fixed, fixed-exponent=-3, drop-exponent=true}
    \begin{tabular}{lcc}
        \toprule
        Method                   & Cart-Pole Swing-Up$^{\text{a}}$             & Truck Backer-Upper$^{\text{a}}$  \\
        \midrule
        \ac{MPPI}                & $\num{0.0571} \pm \num{0.0084}$ & $\num{0.0107} \pm \num{0.0004}$                    \\
        \ac{MPPI Iterative}      & $\num{0.1666} \pm \num{0.0229}$ & $\num{0.0357} \pm \num{0.0020}$                    \\
        \ac{dsMPPI} Permut.      & $\num{0.1675} \pm \num{0.0274}$ & $\num{0.0374} \pm \num{0.0026}$                    \\
        \ac{dsMPPI} Multi-Iter.  & $\num{0.1667} \pm \num{0.0255}$ & $\num{0.0373} \pm \num{0.0027}$                    \\
        \ac{dsCEM} Permut.       & $\num{0.1661} \pm \num{0.0239}$ & $\num{0.0356} \pm \num{0.0028}$                    \\
        \ac{dsCEM} Multi-Iter.   & $\num{0.1670} \pm \num{0.0242}$ & $\num{0.0358} \pm \num{0.0032}$                    \\
        \bottomrule
        \multicolumn{3}{l}{\footnotesize 
        \begin{tabular}[c]{@{}l@{}}
            $^{\text{a}}$mean $\pm$ standard deviation in \si{\milli\second} evaluated over 100 runs 
            \\on a single core of an Intel Xeon Platinum 8358 processor.
            \\
        \end{tabular}
        } \\
    \end{tabular}
    }
\end{table}

%% file: sections/06_conclusion.tex
\section{Conclusion}

In this paper, we proposed \ac{dsMPPI}, a novel sampling-based \ac{MPC} algorithm that integrates deterministic sampling with the exponential weighting scheme of \ac{MPPI}.
Additionally, we introduced a permutation-based variation scheme to enhance exploration with deterministic samples.
Our extensive simulations on two nonlinear benchmark tasks demonstrate that \ac{dsMPPI} effectively leverages the strengths of deterministic sampling and exponential weighting.
It achieves significantly smoother control inputs compared to standard \ac{MPPI} and \ac{dsCEM}, while maintaining competitive cumulative costs.
Furthermore, the proposed permutation scheme was shown to improve performance compared to the multi-iteration approach from \ac{dsCEM}.

These characteristics make \ac{dsMPPI} particularly promising for real-world applications.
By generating smoother trajectories that do not require post-hoc filtering, \ac{dsMPPI} can reduce mechanical stress and extend the lifespan of hardware components.
Crucially, the proposed method incurs no additional online computational overhead compared to random sampling, making it a compelling upgrade for existing sampling-based \ac{MPC} implementations.


%% file: bib/literature.bib
@book{rubinsteinCrossEntropy2004,
  title = {The Cross-Entropy Method},
  author = {Rubinstein, Reuven Y. and Kroese, Dirk P.},
  year = 2004,
  series = {Information {{Science}} and {{Statistics}}},
  publisher = {Springer New York},
  _address = {New York, NY}
}

@inproceedings{chuaDeepReinforcementLearning2018,
  title = {Deep reinforcement learning in a handful of trials using probabilistic dynamics models},
  _booktitle = {Proceedings of the 32nd {{International Conference}} on {{Neural Information Processing Systems}}},
  booktitle = {Proc. 32nd Int. Conf. Neural Inf. Process. Syst. (NeurIPS)},
  author = {Chua, Kurtland and Calandra, Roberto and McAllister, Rowan and Levine, Sergey},
  year = {2018},
  _series = {{{NIPS}}'18},
  pages = {4759--4770},
  _publisher = {Curran Associates Inc.},
  _address = {Red Hook, NY, USA}
}

@inproceedings{hafnerLearning2019,
  title = {Learning latent dynamics for planning from pixels},
  _booktitle = {Proceedings of the 36th {{International Conference}} on {{Machine Learning}}},
  booktitle = {Proc. 36th Int. Conf. Mach. Learn. (ICML)},
  author = {Hafner, Danijar and Lillicrap, Timothy and Fischer, Ian and Villegas, Ruben and Ha, David and Lee, Honglak and Davidson, James},
  _editor = {Chaudhuri, Kamalika and Salakhutdinov, Ruslan},
  year = {2019},
  month = jun,
  _series = {Proceedings of {{Machine Learning Research}}},
  _volume = {97},
  pages = {2555--2565},
  _publisher = {PMLR}
}

@inproceedings{pinneriSampleefficientCrossentropyMethod2021,
  title = {Sample-efficient cross-entropy method for real-time planning},
  _booktitle = {Proceedings of the 2020 {{Conference}} on {{Robot Learning}}},
  booktitle = {Proc. Conf. Robot Learn. (CoRL)},
  author = {Pinneri, Cristina and Sawant, Shambhuraj and Blaes, Sebastian and Achterhold, Jan and Stueckler, Joerg and Rolinek, Michal and Martius, Georg},
  year = {2020},
  month = nov,
  _volume = {155},
  pages = {1049--1065}
}

@inproceedings{eberhard2023pink,
  title = {Pink noise is all you need: {{Colored}} noise exploration in deep reinforcement learning},
  _booktitle = {In {{Proceedings}} of the eleventh {{International Conference}} on {{Learning Representations}}},
  booktitle = {Proc. 11th Int. Conf. Learn. Represent. (ICLR)},
  author = {Eberhard, Onno and Hollenstein, Jakob and Pinneri, Cristina and Martius, Georg},
  year = 2023
}

@article{williamsInformationtheoretic2018,
  title = {Information-theoretic model predictive control: {{Theory}} and applications to autonomous driving},
  shorttitle = {Information-theoretic model predictive control},
  author = {Williams, Grady and Drews, Paul and Goldfain, Brian and Rehg, James M. and Theodorou, Evangelos A.},
  year = 2018,
  month = dec,
  _journal = {IEEE Transactions on Robotics},
  journal = {IEEE Trans. Robot.},
  volume = {34},
  number = {6},
  pages = {1603--1622},
  urldate = {2024-10-08}
}

@article{pezzatoSamplingbasedModelPredictive2025,
  title = {Sampling-based model predictive control leveraging parallelizable physics simulations},
  author = {Pezzato, Corrado and Salmi, Chadi and Trevisan, Elia and Spahn, Max and {Alonso-Mora}, Javier and Hern{\'a}ndez Corbato, Carlos},
  year = 2025,
  _journal = {IEEE Robotics and Automation Letters},
  journal = {IEEE Robot. Automat. Lett.},
  volume = {10},
  number = {3},
  pages = {2750--2757}
}

@inproceedings{bhardwajSTORMIntegratedFramework2022,
  title = {{{STORM}}: {{An}} integrated framework for fast joint-space model-predictive control for reactive manipulation},
  _booktitle = {Proceedings of the 5th {{Conference}} on {{Robot Learning}}},
  booktitle = {Proc. 5th Conf. Robot Learn. (CoRL)},
  author = {Bhardwaj, Mohak and Sundaralingam, Balakumar and Mousavian, Arsalan and Ratliff, Nathan D. and Fox, Dieter and Ramos, Fabio and Boots, Byron},
  _editor = {Faust, Aleksandra and Hsu, David and Neumann, Gerhard},
  year = 2022,
  month = nov,
  _series = {Proceedings of {{Machine Learning Research}}},
  _volume = {164},
  pages = {750--759},
  _publisher = {PMLR}
}

@inproceedings{CDC09_HanebeckHuber,
 address = {Shanghai, China},
 author = {Uwe D. Hanebeck and Marco F. Huber and Vesa Klumpp},
 _booktitle = {Proceedings of the 2009 IEEE Conference on Decision and Control (CDC 2009)},
 booktitle = {Proc. 2009 IEEE Conf. Decis. Control (CDC)},
 _doi = {10.1109/CDC.2009.5400649},
 month = {December},
 _pdf = {CDC09_Hanebeck.pdf},
 title = {Dirac mixture approximation of multivariate {{Gaussian}} densities},
 _url = {http://ieeexplore.ieee.org/document/5400649/},
 year = {2009}
}

@book{kay1993fundamentals,
  title = {Fundamentals of {{Statistical Signal Processing}}: {{Estimation Theory}}},
  author = {Kay, Steven M},
  year = {1993},
  publisher = {Prentice-Hall, Inc.}
}

@inproceedings{schoenauerNeurogeneticTruckBackerupper1994,
  title = {Neuro-genetic truck backer-upper controller},
  _booktitle = {Proceedings of the first {{IEEE Conference}} on {{Evolutionary Computation}}. {{IEEE World Congress}} on {{Computational Intelligence}}},
  booktitle = {Proc. 1st {IEEE} Conf. Evol. Comput., {IEEE} World Congr. Comput. Intell.},
  author = {Schoenauer, M. and Ronald, E.},
  year = 1994,
  pages = {720-723}
}

@inproceedings{arXiv25_Walker,
 _author = {Markus Walker and Daniel Frisch and Uwe D. Hanebeck},
 _doi = {10.48550/arXiv.2510.05706},
 _journal = {arXiv preprint: 2510.05706},
 _month = {October},
 title = {Sample-Efficient and Smooth Cross-Entropy Method Model Predictive Control Using Deterministic Samples},
 _url = {https://arxiv.org/abs/2510.05706},
 _year = {2025},
 _booktitle = {Proceedings of the 2026 {{American Control Conference}} ({{ACC}} 2026)},
 booktitle = {Proc. 2026 {A}m. {C}ontrol {C}onf. ({ACC})},
  author = {Walker, Markus and Frisch, Daniel and Hanebeck, Uwe D.},
  year = 2026,
  month = may,
  pages = {1--8},
  address = {New Orleans, LA, USA}
}
